\documentclass[acmtog]{acmart}

\setcopyright{none}

\usepackage{booktabs} 

\usepackage[ruled]{algorithm2e} 

\SetAlFnt{\small}
\SetAlCapFnt{\small}
\SetAlCapNameFnt{\small}
\SetAlCapHSkip{0pt}
\IncMargin{-\parindent}

\usepackage{amsmath}
\usepackage{amsfonts}
\usepackage{tikz-cd}
\usetikzlibrary{cd} 

\DeclareMathOperator*{\argmax}{\arg\!\max}

\newcommand{\com}		[1]{{}} 

\newcommand{\meshr}{{M}}				
\newcommand{\meshc}{{M^{\prime}}	}		
\newcommand{\VRMax}{{\tilde{M}}}			
\newcommand{\VCMax}{{\tilde{M}^{\prime}}}	
\newcommand{\vr}[1]{{\mathbf{v}_{#1}}}				
\newcommand{\vc}[1]{{\mathbf{v}^{\prime}_{#1}}}		
\newcommand{\msize}{{m}}				

\newcommand{\cager}{{C}	}				
\newcommand{\cagec}{{C^{\prime}}}			
\newcommand{\vcr}[1]{{\mathbf{c}_{#1}}}			
\newcommand{\vcc}[1]{{\mathbf{c}^{\prime}_{#1}}}		
\newcommand{\csize}{{c}}					

\newcommand{\SW}{{\Omega}}				
\newcommand{\SWi}[1]{{\SW_{#1}}}			
\newcommand{\sw}[2]{{\omega_{#1,#2}}}		

\newcommand{\CW}{{\Phi}}				
\newcommand{\CWMax}{{\tilde{\Phi}}}		
\newcommand{\cw}[2]{{\phi_{#1,#2}}}		

\newcommand{\skelr}{{S}}					
\newcommand{\skelc}{{S^{\prime}}}			
\newcommand{\TS}{{\mathcal{T}}}			
\newcommand{\TSi}[1]{{\TS_{#1}}}			
\newcommand{\Ti}[1]{{\mathbf{T}_{#1}}}		
\newcommand{\Tti}[1]{{\mathbf{T}^t_{#1}}}	
\newcommand{\Tri}[1]{{\mathbf{T}^r_{#1}}}	
\newcommand{\ssize}{{s}}					

\newcommand{\JC}{{\Psi}}					
\newcommand{\jc}[2]{{\psi_{#1,#2}}}			

\begin{document}
\title{Real-time Deformation with Coupled Cages and Skeletons}

\author{Fabrizio Corda}
\affiliation{%
  \institution{Università di Cagliari}
  \state{Italy}
}
\email{cordafab@gmail.com}

\author{Jean-Marc Thiery}
\affiliation{%
	\institution{Télécom ParisTech}
	\state{France}
}
\email{jean-marc.thiery@telecom-paristech.fr}

\author{Marco Livesu}
\affiliation{%
	\institution{CNR IMATI}
	\state{Italy}
}
\email{marco.livesu@gmail.com}

\author{Enrico Puppo}
\affiliation{%
	\institution{Università di Genova}
	\state{Italy}
}
\email{enrico.puppo@unige.it}

\author{Tamy Boubekeur}
\affiliation{%
	\institution{Télécom ParisTech}
	\state{France}
}
\email{tamy.boubekeur@telecom-paristech.fr}

\author{Riccardo Scateni}
\affiliation{%
	\institution{Università di Cagliari}
	\state{Italy}
}
\email{riccardo@unica.it}

%


\begin{teaserfigure}
\centering
\includegraphics[width=\linewidth]{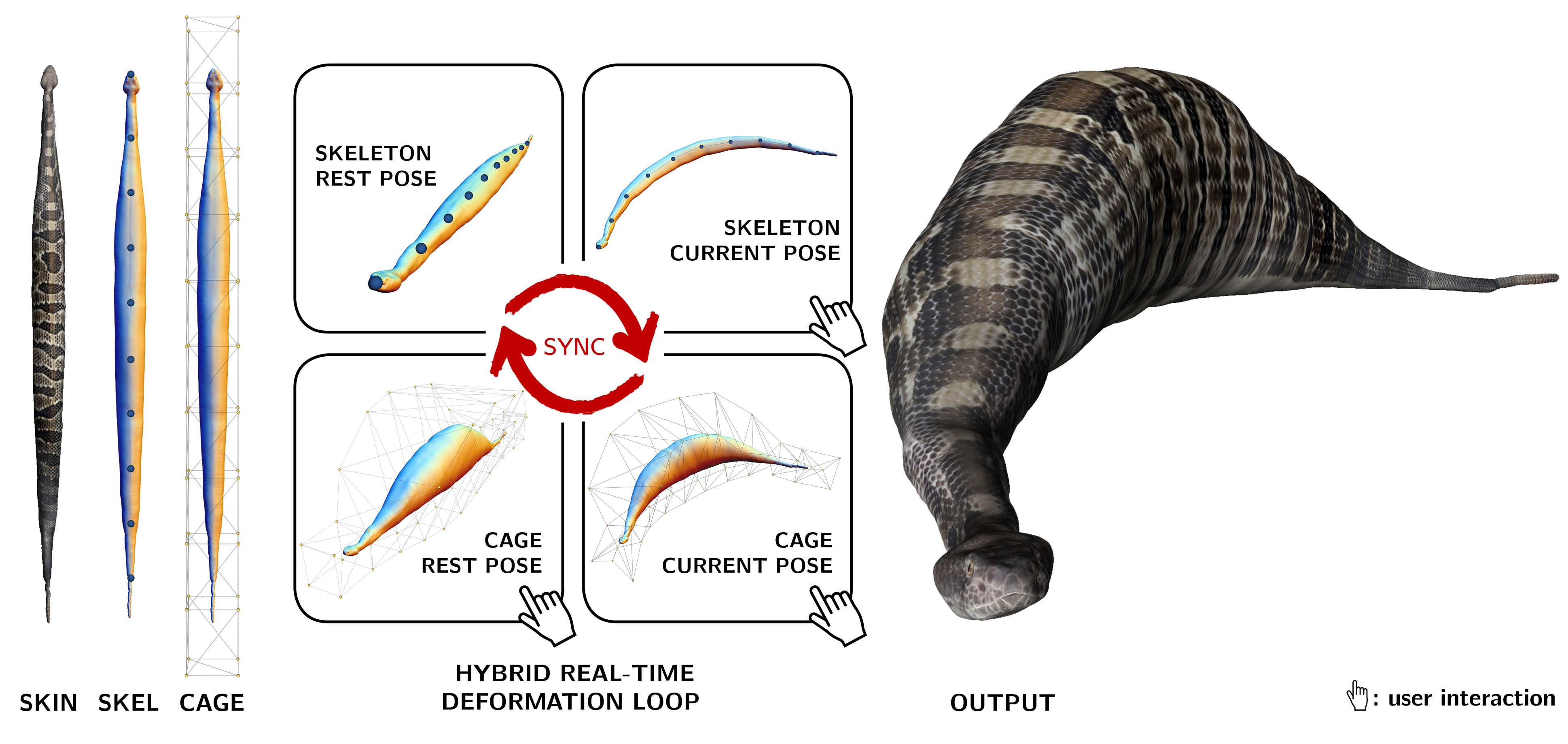}
\caption{Given an input 3D shape equipped with its deformation skeleton and cage (left), our framework seamlessly combines both structures, merging their associated deformation spaces. We achieve this result by using rest and current pose both for the skeleton and the cage (middle). User can deform the shape with any interlaced combination of the two control structures,
editing the skeleton in the current pose, or the cage both in the current and the rest pose. Our novel operators automatically maintain all these entities in sync in real-time.}
\label{fig:teaser}
\end{teaserfigure}

\begin{abstract}
Skeleton-based and cage-based deformation techniques represent the two most popular approaches to control real-time deformations of digital shapes and are, to a vast extent, complementary to one another.
Despite their complementary roles, high-end modeling packages do not allow for seamless integration of such control structures, thus inducing a considerable 
burden on the user 
to maintain them synchronized.
In this paper, we propose a framework that seamlessly combines rigging skeletons and deformation cages, granting artists with a real-time deformation system that operates using any smooth combination of the two approaches.
By coupling the deformation spaces of cages and skeletons, we access a much larger space, containing poses that are impossible to obtain by acting solely on a skeleton or a cage.
Our method is oblivious to the specific techniques used to perform skinning and cage-based deformation, securing it compatible with pre-existing tools. We demonstrate the usefulness of our hybrid approach on a variety of examples.
\end{abstract}

%
%


%
%


\maketitle

\section{Introduction}
\label{sec:intro}
Interactive shape deformation is a fundamental building block in 3D modeling and many other applications.
The various techniques available in the literature rely on simplified control structures: the user interacts with them, and the changes smoothly transfer to the high resolution controlled object (the \emph{skin}) through a set of weights, which establish a relation between each point of the model and the handles of the control structure~\cite{jacobson2014skinning}.
The two most widely used controllers for real-time deformation, namely \emph{skeletons} and \emph{cages}, support complementary tasks: skeleton-based techniques are adequate to control rigid parts and pose articulated bodies;
conversely, cage-based methods are best for smooth volumetric deformations.
Each control structure becomes unwieldy and overly complicated to use where the other excels, thus pushing practitioners to use them together on the same skin.

Numerous discussions on how to combine skeletons and cages can be found on specialized forums, blogs and other online resources. 
Despite this interest from the community of practitioners, the problem of keeping them in sync and finding consensus between the deformations they induce is still open. 
To this end, an important issue 
is that there exists a plethora of alternative skeleton and cage techniques, which are 
already implemented in almost any available deformation software. 
Therefore, to promote a seamless integration of hybrid deformation approaches into well established software packages, it becomes crucial to 
guarantee some sort of flexibility and back-compatibility. 

The complexity in combining skeletons and cages comes from the fact that they achieve deformation in substantially different ways. 
Skeleton-based techniques perform deformations that are \emph{relative} to a particular pose of the skin, often called the \emph{rest} pose: 
the deformed skin is always a combination of its shape in the rest pose with the current position of the skeleton.
Conversely, cage-based deformation follows an \emph{absolute} approach: the current position of the control cage entirely determines the skin it envelopes, with no particular reference pose existing.

Previous attempts
to combine skeletons and cages fall short, either because they 
are not general enough, or because they break back-compatibility.
Blender~\shortcite{blender} allows to link a control cage to a skeleton and move the cage through it. 
The communication is only mono-directional: edits performed on the cage do not reflect on the skeleton, thus requiring complex manual edits to reposition the centers of rotation of each bone. Jacobson and colleagues~\shortcite{jacobson2011bounded} combined skeleton, cage, and point handles into a unified deformation meta-structure, but impose the simultaneous definition of all the handles and relative weights, 
and do not keep the sync between them. Moreover, their system implements a customized pipeline, which is not compatible with standard techniques.

\begin{figure}
\centering
\includegraphics[width=.9\linewidth]{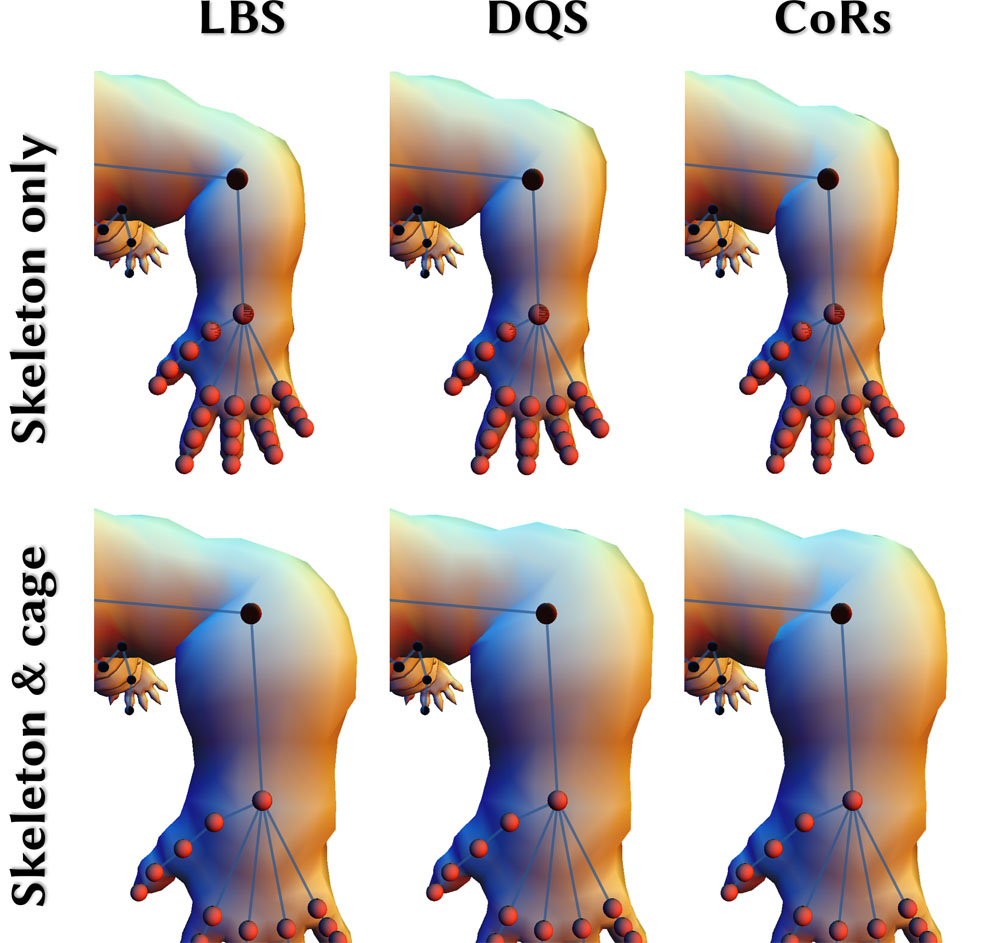}
\caption{Results obtained with various alternative skinning methods implemented in our framework. The top row shows deformations obtained with a skeleton edit only (a 90 degrees rotation of the elbow). The bottom row shows results with an additional cage edit (a uniform stretch of the arm). Our joints and CoRs dynamic repositioning method handles both transformations in a natural manner.}
\label{fig:compSkinning}
\end{figure}

We propose a hybrid deformation paradigm that seamlessly combines skeletons and cages, providing a real-time framework where the user can operate using any interlaced combination of the two control structures. Our method is compatible with classical skeleton-based and cage-based deformation techniques: it just acts as middleware to reach consensus between the two control structures.
In particular, skeleton deformations can be transferred to the skin using the popular LBS~\cite{magnenat1989joint}, DQS~\cite{kavan2008geometric}, or any alternative approach (Figure~\ref{fig:compSkinning}). Linking weights can be either the result of an automatic computation (e.g.~\cite{baran2007automatic}) or hand painted by a digital artist, as it often happens in the industry. Similarly, cage-based deformation admits the use of any type of generalized barycentric coordinates that produce deformation through a linear relation between the cage handles and the vertices of the skin~\cite{nieto2013cage}.
To achieve consensus between the two control structures we retain the relative nature of skeletons
and extend it to the cages,
which therefore exist both in the rest and the current pose (Figure~\ref{fig:teaser}). At modeling time the user guides a deformation by acting on one controller, and the system automatically updates all the other poses accordingly. 
Our approach makes it possible to edit shapes both in current and rest poses, and occurs in real-time, with negligible overhead to classical skeleton- and cage-based deformation workload. Figure~\ref{fig:interactions} provides an overview of our framework, depicting the paths of user interactions.


\paragraph*{Contribution.} Our main contribution is a deformation system that combines 
the deformation spaces of skeletons and cages, revealing a much larger deformation space which contains configurations that cannot be achieved using solely a skeleton or a cage. 
From a technical point of view, our contribution consists in a collection of synchronization operators that maintain the pose set up-to-date during the editing session in real-time. 
We demonstrate our hybrid deformation approach on a variety of examples, including shape modeling and digital animation (Section~\ref{sec:results}). Our software prototype already implements several skeleton and cage based techniques, and can potentially embrace most existing techniques based on these control structures, thus demonstrating a great flexibility and back-compatibility. 

\section{Related works}
\label{sec:related}
Our approach bridges skeleton-based and cage-based deformations. While works of this kind are quite rare, the literature offers a wide variety of approaches that focus on one deformation paradigm or the other. We refer the reader to the work of Jacobson et al.~\shortcite{jacobson2014skinning} for a recent survey, and focus here on the articles most relevant to our work. 

\paragraph*{Skeletons}
Deformations defined by linearly blending transformation matrices associated to the bones of a control skeleton, also known as Linear Blend Skinning (LBS), appeared long time ago in the literature~\cite{magnenat1989joint} and have become extremely popular ever since. LBS suffers from a number of artifacts, most of which come from singular transformation matrices generated by the linear interpolation of rotations. 
A typical artifact of this kind is the well known \emph{candy wrap}, which arises at the shoulders of a character when big torsions are applied to its arms. Recent literature has proposed more robust ways to combine transformation matrices. In particular,  Kavan and colleagues proposed to blend rotations in the space of Dual Quaternions (DQS), avoiding the generation of candy wrap artifacts at the cost of a minimal overhead in the real-time deformation pipeline \cite{kavan2008geometric}. 
Due to their simplicity and intuitiveness, LBS and DQS are \emph{de facto} standards in skeleton-based deformation.
More recent non-linear skinning methods, such as Stretchable and Twistable bones \cite{jacobson2011stretchable}, Differential Blending \cite{Oztireli13Differential} and Elasticity Inspired Deformers \cite{Kavan-12-EID} introduce more sophisticated techniques that act on the same basic ingredients, namely rigging weights and affine transformations, to define the pose of the skeleton.
Finally, Le and Hodgins~\shortcite{le2016real} introduced an efficient method designed to avoid the respective artifacts of LBS and DQS, through the definition of a per-vertex Center of Rotation (CoR) derived from an analysis of the skinning weights, and resulting from an optimization targeting as-rigid-as-possible deformations.
In our software prototype we incorporated LBS, DQS and CoRs (Figure~\ref{fig:compSkinning}), but our framework can support all previously cited skinning methods, thus promoting a seamless integration with available implementations.
Critical to the previously cited methods is the definition of the so-called skinning weights.
Various automatic methods aim at defining smooth weights~\cite{baran2007automatic,wareham2008bone,jacobson2011bounded,jacobson2012smooth,dionne2013geodesic}
or target as-rigid-as-possible deformations under training~\cite{thiery2018araplbs}.
While automatic methods generally provide satisfactory results on challenging inputs, artists often need to tune the skinning weights as they target specific effects,
and methods allowing for high-level skinning weights editing have been developed recently~\cite{bang2015interactive,bang2018spline}.
Our method is quite general and has no restrictions on the skinning weights used for deformation, thus
allowing for great artistic flexibility.

\paragraph*{Cages}
Cage-based methods derive directly from the Free Form Deformation \cite{sederberg1986free} and allow to deform
a volume bounded by a control mesh. 
Each point within such volume is defined as a weighted sum of cage vertices, hence its position can be efficiently updated each time the cage is deformed by the user. Control weights are generalized barycentric coordinates, and differ to each other mainly for their smoothness and locality.
Several alternatives have been proposed in the literature~\cite{TMB:2018:QMVC,TTB:2013:JHMVC,ju2005mean,joshi2007harmonic,lipman2007gpu,lipman2008green,Zhang:2014:LBC,hormann2008maximum}.
All methods, with the exception of~\cite{lipman2008green,ben2009variational}, compute skin coordinates as a linear combination of cage vertices, and are seamlessly supported by our framework.
Similarly to skeletons, cages can be either automatically computed or manually crafted, thus ensuring a seamless integration with most available implementations. For brevity, 
we do not review methods for cage generation. 
We point the reader to the survey of~\cite{nieto2013cage} for classical literature in the field, and to~\cite{CLMASBP19} and references therein for a list of more recent algorithms.


\paragraph*{Hybrid approaches}
Some methods depart from the classical skele\-ton-based and cage-based deformation paradigms, trying to improve on them on some aspect. Garcia et al.~\shortcite{Garcia:2013:CMM} proposed a hybrid system that seamlessly combines multiple cages and barycentric coordinates. The system is completely devoted to cages only, and does not take into consideration interactions with a skeleton. 
Mukai and Kuriyama~\shortcite{mukai2016efficient} propose the use of automatically generated bone helpers to enrich the space of deformations of linear blend skinning with secondary motions, enabling the animation of muscles and soft tissues. The system focuses on a very specific problem and does not offer the flexibility granted by a real control cage. Being compatible with LBS, their bone helpers could also be incorporated in our framework.
Ju et al.~\shortcite{Ju:2008:RST} combined the use of cages and skeletons to avoid the candy wrap artifacts of LBS. 
Opposite to ours, their system works as an open loop, using the skeleton to pose the cage, and the cage to pose the skin. Similar systems are currently supported by commercial software (e.g. Blender~\shortcite{blender}), but do not really offer the possibility to seamlessly combine deformations defined on the skeleton with others defined on the cage in arbitrary order.
The combination of skeletons and point handles was explored in~\cite{wang2015linear}. 
Jacobson et al.~\shortcite{jacobson2011bounded} proposed a system where skeleton, cage and point handles are all integrated into the same framework. Their system require the simultaneous definition of all structures (see Equation 1 in the original paper), and is based on the use of the same coordinates at all levels, without permitting the use of manually painted weights, or different weights for different handles.
Moreover, the skeleton and the cage are part of the same meta-structure and must be jointly animated: manipulating the skeleton (resp. the cage) will not induce a deformation of the cage (resp. the skeleton), contrary to what we aim at. 
All in all, previous techniques cannot offer the flexibility of our technique, that is, switching seamlessly from one structure to the other while always relying on the optimization framework to update the other structure appropriately.
Finally, combining deformation rigs has been used for a long time in specialized industrial scenarios. In particular, on-surface facial rigs are often superimposed on the skeleton rigs that control the head orientation. However, this scenarios typically come with a fixed prioritization policy e.g., facial rigs defined in the local frame of the face, which itself undergoes a single rigid transform stemming from the neck bone. The scenarios we address is more challenging, as both structures compete to control global and non-rigid deformations that happen simultaneously.

\paragraph*{Exploration of shape space} Our method is loosely related to methods based on the analysis of 3D shapes and subsequent structure aware deformation, which have the main intent to explore the space of shapes similar to a reference mesh (Figure~\ref{fig:airplane}). To this end, our real time deformation tool complements classical methods based on feature curves~\cite{gal2009iwires}, bounding envelopes~\cite{zheng2011component} or semantic handles~\cite{yumer2015semantic}, without exposing any significant technical overlap.

\section{Background}
We take as input a polygonal mesh $\meshr_0$, together with a skeleton $\skelr_0$ rigged to $\meshr_0$, and a cage $\cager_0$ surrounding $\meshr_0$. 
We set our working structures at rest pose by initializing $\meshr\equiv\meshr_0$, $\skelr\equiv\skelr_0$ and $\cager\equiv\cager_0$; note that 
$\meshr$, $\cager$ and $\skelr$ can evolve because of editing, as discussed in the next section. 
Our method uses only the set of vertices 
of meshes and is oblivious of their connectivity.
For this reason, whenever no ambiguity arises, we will overload the same symbol to denote both a mesh and its set of vertices. 

\paragraph{Skeleton-based deformation}
As customary, instead of representing the skeleton $\skelr$ explicitly, we consider the set of rigid transformations $\TS=\{\Ti{1},\ldots,\Ti{\ssize}\}$ that determine a given pose. Each transformation $\Ti{j}$, is associated to the $j$-th bone of $\skelr$ and represents a rotation around one of its endpoints, which affects all the sub-skeleton below the given joint in the bones hierarchy. 
At rest pose, all transformations are set to identity.
The skeleton $\skelr$ is rigged to $\meshr$ through a $\msize \times \ssize$ (sparse) matrix of weights $\SW$, where each entry $\sw{i}{j}$ defines the influence of the $j$-th bone of $\skelr$ on $i$-th vertex of $\meshr$.  

A general skeleton-based deformation (a.k.a.\ skinning) has the following form
\begin{equation}
\label{eq:skinning}
\meshc = F(\TS,\SW,\meshr)
\end{equation}
where  $\meshc$ is the deformed (current) mesh.
For instance, LBS is encoded vertex-wise by 
\begin{equation}
\label{eq:LBS}
\vc{i}=\sum_{j=1}^{\ssize} \sw{i}{j} \Ti{j} \vr{i},
\end{equation}
which is often presented in the following form where the linear part applied to the vertex is separated from the translation part
\begin{equation}
\label{eq:lbs_with_rotations_and_translations_separated}
\vc{i}= (\sum_{j=1}^{\ssize} \sw{i}{j} \Tri{j}) \cdot \vr{i}  +  (\sum_{j=1}^{\ssize} \sw{i}{j} \Tti{j}),
\end{equation}
while DQS is encoded similarly
\begin{equation}
\label{eq:DQS}
\vc{i}=\mbox{\sc DQblend}(\TSi{i},\SWi{i})\ \vr{i},
\end{equation}
where {\sc DQblend} is the proper function to blend transformations represented via dual quaternions,
while $\TSi{i}$ and $\SWi{i}$ denote the $i$-th rows of $\TS$ and $\SW$, respectively.

The recently introduced CoR method~\cite{le2016real} is slightly different from other methods, in the sense that it makes use of an additional parameter derived from a cross analysis of the mesh geometry and the skinning weights: a per-vertex Center of Rotation (CoR).
This CoR $p_i$ associated with vertex $i$ is computed as
\begin{equation}
\label{eq:CORdefinition}
 p_i := \int\limits_{x\in M}{ \delta( \omega_{i,\cdot} , \omega_{x,\cdot} ) \, x\, dx }\;  / 
       \int\limits_{x\in M}{ \delta( \omega_{i,\cdot} , \omega_{x,\cdot} ) dx } ,
\end{equation}
$\delta( \omega_{i,\cdot} , \omega_{x,\cdot} )$ denoting a distance between the sets of weights of vertex $i$ and vertex $x$.

Given the CoR $p_i$ associated with vertex $i$, and computing the linear part $R(i)$ applied to the mesh vertex using quaternion blending of the bone rotations $\{\Tri{j}\}_j$,
the translation applied to the vertex is computed as
$$
t(i) = \sum_{j=1}^{\ssize}\sw{i}{j} (\Tri{j} \cdot p_i  +  \Tti{j})  -  R(i) \cdot p_i.
$$

Since the CoR computation depends only on the skinning weights, vertices with similar skinning weights have similar CoRs and are therefore transformed by the same rigid transformation.

Overall, the (run time) deformation of vertex $i$ by the CoRs method can be summarized as
\begin{equation}
\label{eq:CORSdeformation}
\begin{cases}
\vc{i} = R(i) \cdot \vr{i} + t(i) &\text{ (deformation)} \\
R(i) = \text{\sc DQBlendRot}(\{\omega_{i,j},\Tri{j}\}_j) &\text{ (linear part)} \\
t(i) = \tilde{p}_i  -  R(i) \cdot p_i &\text{ (translation part)} \\
\tilde{p}_i = \sum_{j=1}^{\ssize} \sw{i}{j} (\Tri{j} \cdot p_i +\Tti{j}) &\text{ (transformed CoR)} \\
\end{cases}
\end{equation}
where {\sc DQBlendRot} reurns the linear part of the matrix {\sc DQBlend} as defined above.

Our method is compliant with all skinning methods mentioned in Section~\ref{sec:related}, and any which requires only the rest pose location of the joints of the skeleton (or centers of rotation derived from the rest pose mesh as done in the CoRs method).

\paragraph{Cage-based deformation}
The cage $\cager$ controls $\meshr$ via a $\msize \times \csize$ (sparse) matrix of barycentric coordinates $\CW$, where entry $\cw{i}{k}$ is the barycentric coordinate of mesh vertex $\vr{i}$ with respect to cage vertex $\vcr{k}$. 
We require deformation to be given vertex-wise by the standard linear equation
\begin{equation}
\label{eq:caging}
\vc{i}=\sum_{k=1}^{\csize} \cw{i}{k} \vcc{j}
\end{equation}
where $\vc{i}$ represents the (deformed) position of vertex $\vr{i}$ of $\meshr$ when the cage vertices are set at (edited) positions $\vcc{j}$.
This equation is compliant with all barycentric coordinates reviewed in Section \ref{sec:related}, except the Geeen coordinates~\cite{lipman2008green} and the Variational Harmonic Maps~\cite{ben2009variational}, which require also face normals that set a non-linear relation between the cage and the skin. Indeed, our method is not compliant with the latter two techniques.
Note that Equation \ref{eq:caging} does not refer to any rest position for either the cage or the skin. 
By convention, we assume that we have a rest pose for the cage $\cager$, when its vertices are at positions 
where equality 
$\meshr=\CW \cager$ holds. 
Namely, the rest pose of the cage induces the rest pose of the skin. 
In fact, this is the usual setting from which the barycentirc coordinates are obtained.

%

%


\begin{figure*}
\centering
\includegraphics[width=.95\linewidth]{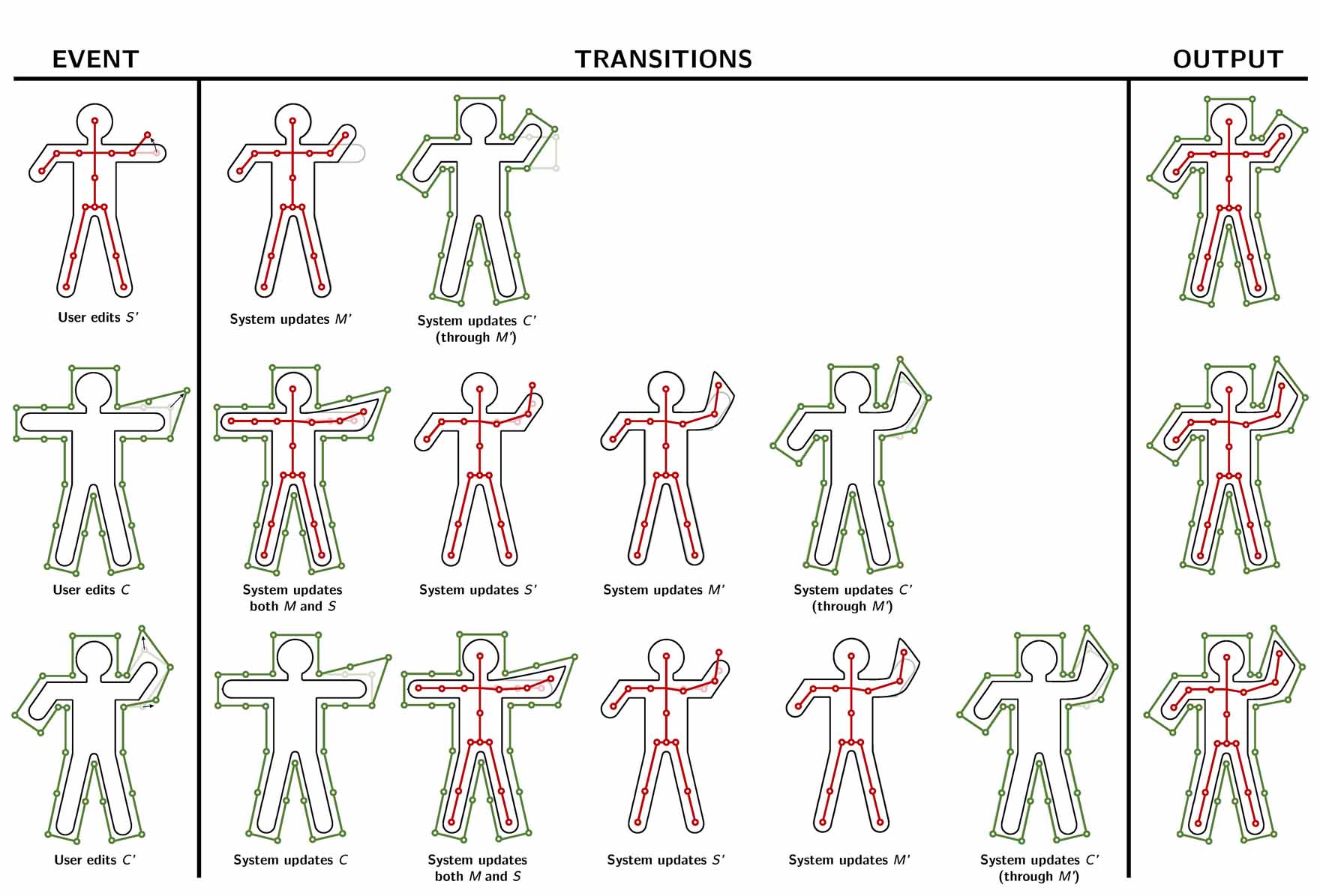}
\caption{Interactions with our hybrid deformation system: the user can edit the skeleton in the current pose ($\skelc$) or the cage, both in the rest ($\cager$) and the current ($\cagec$) pose. The diagram illustrates the chain of reactions the automatically update the system, maintaining the sync among the various entities involved in the deformation. All interactions occur in real-time.}
\label{fig:interactions}
\end{figure*}

\section{Method}
\label{sec:method}
In the following, we will refer to six structures, namely: the skin $\meshr$, the skeleton $\skelr$ and the cage $\cager$ at rest pose, together with their deformed counterparts $\meshc$, $\skelc$ and $\cagec$ at current pose.
Our method aims at reaching consensus among these structures upon any interleaved sequence of skeleton-based and cage-based deformations. Our system makes it possible to edit any structure, assuming the sets of skinning weights $\SW$ and barycentric coordinates $\CW$ remain constant.
Note that $\meshr$ is induced by $\skelr$ via skinning when $\TS$ contains just identity transformations; and $\meshr$ is induced by $\cager$ via barycentric coordinates: these invariants set the consensus at rest pose and will be maintained throughout. 
Unfortunately, if a mesh $\meshc$ is induced via skinning by a skeleton $\skelc$, there may not exist a cage $\cagec$ that induces $\meshc$, and vice-versa. We address this issue by synchronizing $\skelc$ and $\cagec$ so that they produce similar skins:
$\meshc$ is obtained through skinning, with a process that incorporates also $\cagec$. 

\paragraph*{User interaction.}
The user can operate on three of the six structures defined above, namely the user can edit:
\begin{itemize}
\item the current pose $\skelc$ of the skeleton;

\item the cage $\cager$ at rest pose;

\item the cage $\cagec$ at current pose.
\end{itemize}

Direct editing of either the skin (resp. skeleton) at rest pose is not considered, as this operation is inherent to modeling (resp. rigging) the input shape; it would be straightforward to include them in our framework, though.
Note that the roles of the skeleton and the cage are not fully symmetric: while the former can modify only the current pose of the skin, the latter can modify both current and rest poses.
From the user point of view, editing the rest pose means adapting a different skin to the rigging, which remains untouched.

When executed, the three user interactions outlined above trigger a set of sync operations, automatically handled by the system. Specifically:
\begin{itemize}
\item if $\skelc$ is edited, then both $\meshc$ and $\cagec$ will follow, while no change will occur to any structure in the rest pose (Figure~\ref{fig:interactions}, top);

\item if $\cager$ is edited, then $\meshr$ will follow,  $\skelr$ will be adjusted to the new skin and the current set of transformations $\TS$ will be adjusted accordingly, 
inducing a new current pose $\skelc$ and a new skin $\meshc$ -- the current cage $\cagec$ will be also updated (Figure~\ref{fig:interactions}, middle);

\item if $\cagec$ is edited, we have the most complicated situation:
changes to $\cagec$ will be reflected to $\cager$; while all other modifications will occur as in the previous case up to $\cagec$ itself, in a closed loop (Figure~\ref{fig:interactions}, bottom).
\end{itemize}
In order to get a consistent result in all cases, we must synchronize the different structures, as explained in the following. 

\begin{figure}[ht]
	\centering
\begin{tikzcd}
	&															& \arrow[hook, two heads]{rd}{\mbox{\small\sc UI}} \\
	&\skelr \arrow[dotted]{ld}[above]{\SW} \arrow{rr}[above]{\TS}				&	& \skelc \arrow{r}[above]{\SW} \arrow{dd}{\mbox{\small\sc CageUp}}	& \meshc\\
\meshr\\
	&\cager \arrow{ul}[below]{\CW}	\arrow{uu}[right]{\mbox{\small\sc SkelUp}}		&	& \cagec \arrow{ll}{\mbox{\small\sc CageRev}} \\	
\arrow[hook, two heads]{ru}[below]{\mbox{\small\sc UI}} & 					& \arrow[hook, two heads]{ru}[below]{\mbox{\small\sc UI}}
\end{tikzcd}
\caption{A diagram of our method showing transitions among the different structures; symbols attached to arrows denote the parameters or algorithms used to effect such transitions; hooked arrows marked {\sc UI} denote User Interaction. 
	The dotted arrow represents rigging at rest pose and is just symbolic.
} 
\label{fig:diagram}
\end{figure}
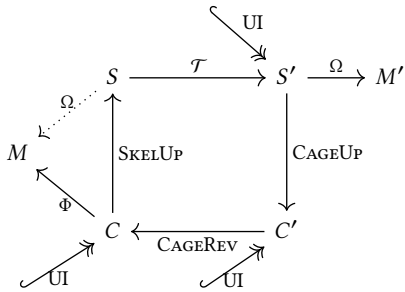

\paragraph*{Synchronization.}
We achieve synchronization among the six structures in the two poses 
by introducing a set of \emph{transitions}, which reflect editing among them and are summarized in Figure~\ref{fig:diagram}:
\begin{itemize}
\item $\skelr\rightarrow \skelc\rightarrow \meshc$ is the standard skinning from Equation \ref{eq:skinning};
\item $\cager \rightarrow \meshr$ 
is the standard application of barycentric coordinates from Equation \ref{eq:caging}; 
\item $\cager \rightarrow \skelr$ adjusts the centers of rotations of $\skelr$ on a modified skin $\meshr$ at rest pose -- we call it the \emph{skeleton updater} {\sc (SkelUp)};
\item $\skelc \rightarrow \cagec$ finds a consensus between the current poses of the skeleton and the cage, -- we call it the \emph{cage updater} {\sc (CageUp)}; 
\item $\cagec \rightarrow \cager$ reflects the direct editing on the current pose to the cage at rest pose -- we call it the \emph{Reverse cage deformer} {\sc (CageRev)}.
We must guarantee that the current cage $\cagec$ determined from editing by the user coincides with the current cage resulting from the transition $\cager\rightarrow\skelr\rightarrow\skelc\rightarrow\cagec$, as induced by the rest cage $\cager$ modified by $\cagec$. In order to achieve this result, we must set this transition in a proper way.
\end{itemize}
No other direct transitions are considered.
For instance, there is no direct transition from $\cager$ to $\cagec$.
This is a specific design choice, which determines the 
clockwise central cycle
$\cager\rightarrow\skelr\rightarrow\skelc\rightarrow\cagec\rightarrow\cager$ in Figure \ref{fig:diagram}:
no matter where interaction starts, we propagate its effects through this cycle to maintain all four structures synchronized.

The four structures in the central cycle of the diagram are synchronized after each editing operation so that the central cycle remains at a steady state. 
The non-standard operations in the above list are detailed in the following subsections.


\subsection{Cage Updater}
\label{sec_cage_updater}
The transition $\skelc\rightarrow\cagec$ corresponds to the algorithm denoted {\sc CageUp} in Figure \ref{fig:diagram}. When the skeleton $\skelc$ is modified, we skin $\meshr$ to obtain $\meshc$ and update $\cagec$ accordingly. 
More precisely, we seek a cage $\cagec$ that generates a skin as close as possible to $\meshc$, according to the (static) barycentric coordinates $\CW$.


As already observed in Section~\ref{sec:method} this problem may (and in general does not) admit an exact solution. To avoid solving a least squares problem of the size of $\meshc$, we apply the \emph{MaxVol} relaxation proposed in \cite{Thiery:2012:CCR}.
During pre-processing we extract a subset $\VRMax$ of vertices of $\meshr$ with the same cardinality of the vertices of $\cager$. Vertices are selected so as to result in a matrix of coordinates with the highest volume. 
Then, we consider the corresponding set $\VCMax$ in the current mesh $\meshc$, and we solve the linear system
\begin{equation}
\label{eq:cupd}
\CWMax \cagec = \VCMax,
\end{equation}
where $\CWMax$ is the submatrix of 
$\CW$ corresponding to the vertices of $\VRMax$.
Note that this is a square system having same size of $\cagec$ (which is assumed to be much smaller than $\meshc$). The system is invertible and remains fixed throughout; we factorize it once and efficiently solve it with back-substitution in real-time.
Besides performances, as shown in~\cite{Thiery:2012:CCR} the resulting fitted cages are also more stable than the cages fitted to the full geometry using a least squares approach. Also consider that the purpose of the {\sc CageUp} is not to best reconstruct the mesh as a function of the fitted cage, but rather to provide the user with a stable cage that nicely envelopes it and aids interaction, therefore the use of the relaxation is appropriate.



\begin{figure*}
\includegraphics[width=\linewidth]{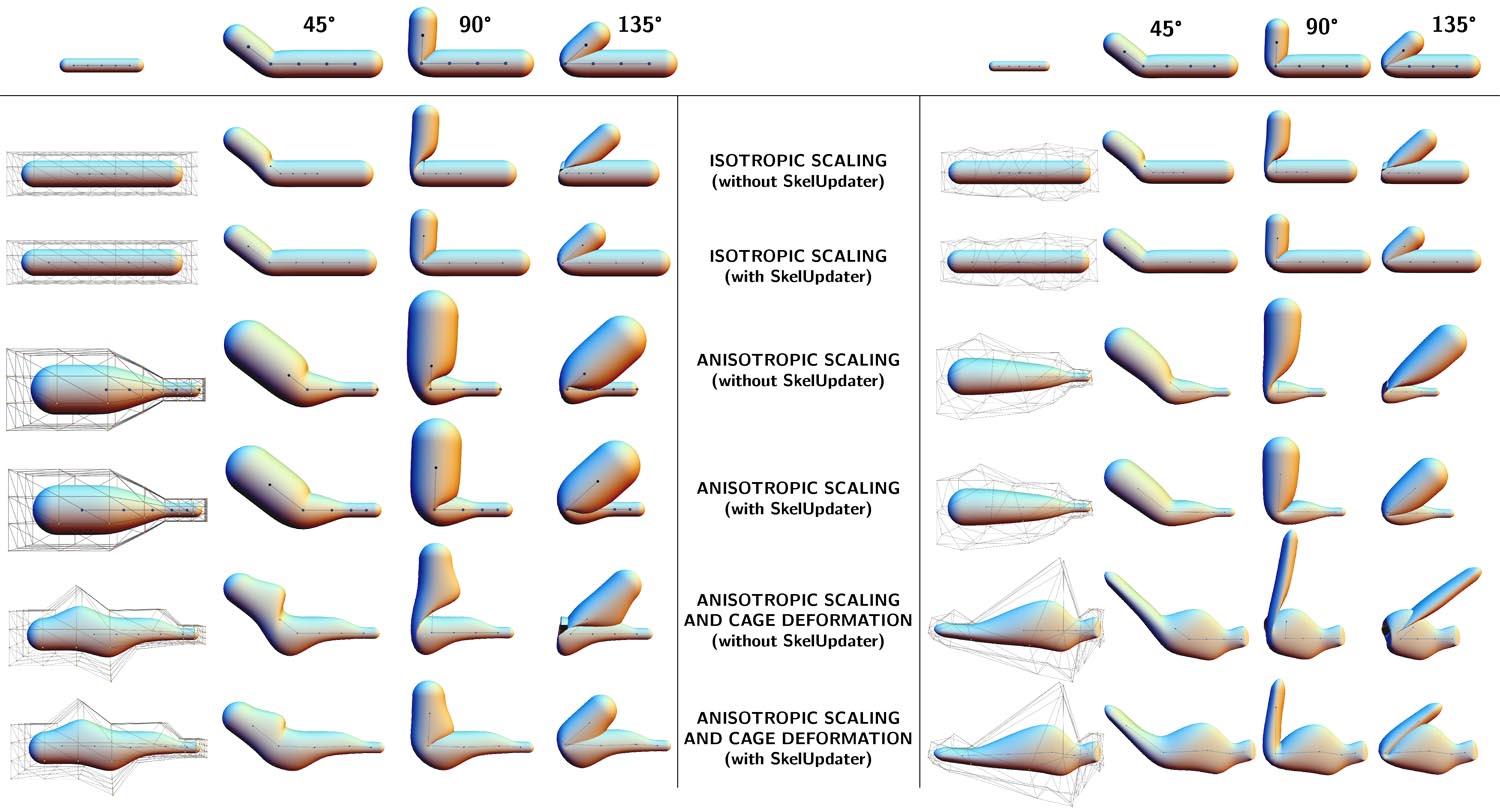}
\caption{A straight bar bent at $45, 90$ and 135 degrees using LBS, on top of which we applied various skeleton edits (isotropic and anisotropic scaling, single handle displacement). When the skeleton updater is disabled, cage edits move the skeleton away from the correct centers of rotation, generating various visual artifacts. Our bone positioning system correctly recovers from all configurations, producing visually plausible deformations. Note that the system is not sensitive to the specific cage being used, and produces valid results both with a regular (left) and irregular (right) cage.}
\label{fig:eval_skelupdater}
\end{figure*}

\subsection{Skeleton Updater}
\label{sec:supd}
The transition $\cager\rightarrow\skelr$ corresponds to the algorithm denoted {\sc SkelUp} in Figure~\ref{fig:diagram}.
When a deformation of $\cager$ is performed -- either directly or as the result of deforming cage $\cagec$ -- we update the skeleton at rest pose and propagate it down the skinning pipeline.
Indeed, when the cage stretches limbs or creates a bulge on the skin, the position of the skeleton joints, which act as centers of rotation during skinning, must be repositioned to avoid artifacts (Figure \ref{fig:eval_skelupdater}).
In other words, we need this to preserve the semantic relation between the skeleton joints and their position relative to the skin.

We address this issue by introducing a new relation between the cage vertices and the position of skeleton joints. This relation is computed once and for all in pre-processing, and allows to express the position of skeleton joints at rest pose as a linear combination of the cage vertices, giving us the ability to readily update/refit the skeleton in real-time. Note that this update is not limited to a simple global registration, but rather can change the local geometry of the skeleton (e.g. stretching/shrinking its bones).

\begin{figure}[ht]
	\centering
	\includegraphics[width=.9\linewidth]{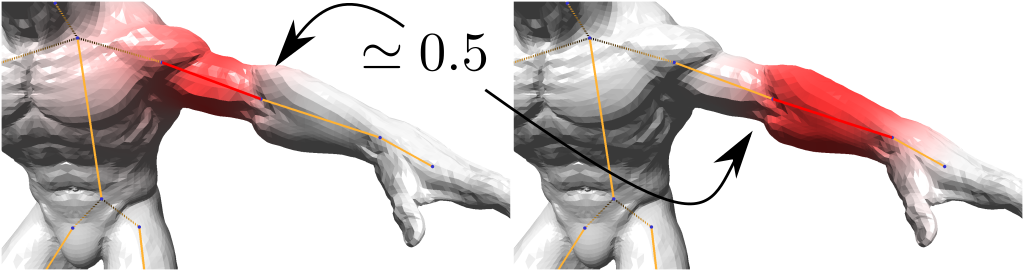}
	\caption{
Weights are typically close to $1$ around the middle of the bones, resulting in highly rigid transformations in these parts of the mesh, while they blend most with the other weights near the articulations.
	}
	\label{fig:lbsweights_localization_function_motivation}
\end{figure}

\paragraph*{Skeleton Updater weights}
All computations in the following are peformed only once in the initial pose and involve $\cager\equiv\cager_0$ and $\skelr\equiv\skelr_0$.
We first identify to which mesh vertices a joint $j$ corresponds to, by defining a (discrete) \emph{joint localization function} $L^{\SW}_{j,\cdot}$ for each joint of $\skelr$.
These functions depend only on the skinning weights $\SW$ and are defined vertex-wise on $\meshr$ as follows
\begin{equation}
\label{eq:joint_localization_function}
L^{\SW}_{j,i} = -1 + \sw{i}{j}^s + (\sum_{k \neq j}{\sw{i}{k}})^s,
\end{equation}
with $s << 1$ (we use $s=0.1$ in our implementation).
Function $L^{\SW}_{j,\cdot}$ takes value $0$ in rigid regions (i.e., for $\sw{i}{j}=1$ and $\sw{i}{j}=0$) and larger values as $\sw{i}{j}$ approaches $0.5$, i.e., near the joint,
where the skinning weights blend the most 
(see Figure~\ref{fig:lbsweights_localization_function_motivation}).

Next, we use our joint localization functions to define barycentric coordinates for the joints positions $\{a_j\}$ w.r.t.\ the cage, and exploit them to transform the joints along with the skin with cage deformation.
Specifically, we first compute Mean Value Coordinates $\{\text{mvc}_{j,i}\}_i$ for the joints rest pose locations $a_j$ w.r.t. the input \emph{mesh},
which we localize around the articulation using the localization function.
Note that the resulting weights $\{\mathbf{L}^{\SW}_{j,i}\}_i := \{\text{mvc}_{j,i}*L^{\SW}_{j,i}\}_i$ are not valid barycentric coordinates at this step.
The joints barycentric coordinates are then defined as
\begin{equation}
\label{eq:joint_weights_wrt_the_cage}
\jc{j}{\cdot } = \mbox{\sc MEC}\left( \sum_{i}{\mathbf{L}^{\SW}_{j,i} \cw{i}{\cdot}} \right),
\end{equation}
where the free index $\cdot$ is varying over the vertices of $\cager$ and $\mbox{\sc MEC}$ denotes the projection of input masses in
$\mathbb{R}^{\csize}$ to the set of valid barycentric coordinates for $a_j$
i.e., verifying linear precision: $a_j = \sum_{k}{\jc{j}{k} c_k}$ and partition of unity: $\sum_{k}{\jc{j}{k}} = 1 ~ \forall j$ -- and closest to the input masses as the output barycentric coordinates maximize the cross entropy, following the strategy introduced by Hormann and Sukumar~\shortcite{hormann2008maximum}
\begin{eqnarray} \label{eq:recap_on_mec}
MEC\left( \{m_k\} \right) := &\argmax\limits_{\{b_k\}} \sum_k{ - b_k ln( b_k / m_k )  } , \\
s.t.
&\begin{cases}
 \sum_k{ b_k . c_k } = a_j \\
 \sum_k{ b_k } = 1 \\
\end{cases}\nonumber
\end{eqnarray}


One may see this construction as deriving barycentric coordinates for the input articulations $A$ w.r.t.\ the input cage $\cager_0$ through the combination of (i) the input cage coordinates, and (ii) the localization function derived from the input skinning weights, which allows us expressing (once fixed) the articulations as a linear combination of the cage vertices, since
\begin{equation} \label{eq:joint_positions_wrt_the_cage}
A = MEC(  \mathbf{L}^{\SW} \cdot \CW  ) \cdot \cager := \JC \cdot \cager
\end{equation}
where $MEC(\cdot)$ is computed here per line $j$ for each joint $j$ with constrained rest-pose $a_j$ independently.
This construction presents several advantages.
In particular, we make no assumption over the set of input coordinates $\CW$, input skinning weights $\SW$, or quality of the input mesh.
Additionally, the construction is intuitive, since one can simply edit the localization function $L^{\SW}_{i,j}$ as an ad-hoc set of weights allowing for the reconstruction of the joint position as a combination of the mesh vertices.
Moreover, it is highly efficient and parallelizable.

\paragraph*{Skeleton joints refitting}
When $\cager$ is deformed, we update the joints of $\skelr$ at positions $\{a_{j}\}$ as a linear combination of cage vertices with
\begin{equation} \label{eq:skRefit}
a_{j} = \sum_{k=1}^{\csize} \jc{j}{k} \vcr{k},
\end{equation}
where the $\jc{j}{k}$ are the barycentric coordinates described earlier.
Note that, after the joints have been relocated, the length and orientation of the bones in the skeleton have been changed; these changes must be reflected on the current skeleton $\skelc$ and, consequently to the $\cagec$ and $\meshc$.
In order to trigger these changes, we update each skinning transformation $\Ti{j}$ of $\TS$ by keeping its rotational component $\Tri{j}$ unchanged and by recomputing its translational component $\Tti{j}$ according to the new joints rest-pose locations.
We do so by simply following the hierarchical structure of the skeleton, updating first all roots, and then processing children in an iterative manner, so as to preserve the iteratively deformed skeleton articulations.
The effect of updating $\TS$, hence $\skelc$, propagates down through standard skinning, and the Cage Updater described previously.



\paragraph*{CoRs repositioning}
As already discussed, the CoRs method makes use of per-vertex centers of rotations $p_i$ for vertices $i$, which are precomputed following Equation~\ref{eq:CORdefinition}. 
Since a manipulation of the cage deforms the rest pose state for the skeletal deformation, we have to reposition the CoRs as well. Fortunately, those are defined as a linear combination of the mesh rest pose vertices $M$.
Rewriting Equation~\ref{eq:CORdefinition} in matrix form as $CoRs = \Phi_{CoRs} \cdot M$,
and using the fact that the rest pose mesh $M$ is expressed as $M = \Phi \cdot C$,
we can precompute the matrix $\Lambda := \Phi_{CoRs} \cdot \Phi$ when computing the centers of rotation, and reposition them at run time using
\begin{equation}
 CoRs = \Lambda \cdot C.
\end{equation}
Doing so, we assume that the area terms in the surface averaging remain similar (see Equation~\ref{eq:CORdefinition}).
In fact, this introduces slight differences between the CoRs we obtain after a cage deformation of the rest pose mesh, and the ones one can obtain when recomputing them from scratch, every time the rest pose mesh is changed.
However, these differences are minor, and do not impact negatively the quality of the resulting deformation (see Figure~\ref{fig:compSkinning}).
In particular, the vertices with similar input skinning weights are still transformed by the same rigid transformation. Lastly, our joints repositioning method is highly compatible in spirit with the CoRs method, as
both motivate the use of the cross analysis of the mesh geometry and the skinning weights in the derivation of optimal pivot positions.


%
\subsection{Reverse Cage Deformer}
\label{sec:rskin}
The transition $\cagec\rightarrow\cager$ corresponds to the algorithm denoted {\sc CageRev} in Figure \ref{fig:diagram}.
%
While the user interacts with $\cagec$, we must reflect any modification $\partial\cagec$ with a corresponding modification 
$\partial\cager$ of the rest cage $\cager$, and such modification must maintain the framework consistent.
To do so, we express the generic modification $\partial\cagec$ as a function of $\partial\cager$ through the sequence 
$\cager\rightarrow\skelr\rightarrow\skelc\rightarrow\cagec$, and we finally reverse this function.
In order to obtain a linear problem and achieve efficiency, we compute the above function by assuming LBS throughout.
We will discuss at the end of the section how to handle other types of skinning methods.

In order to exploit matrix computation, we linearize all our structures.
For the sake of clarity and with abuse of notation for this section only, we use the same symbols as before to denote the linearized structures.
We denote $\cager,\cagec \in \mathbb{R}^{3\csize}$ the vectors stacking the vertices of the rest and the current cage, respectively; 
and we denote $\SW$, $\CW$, $\JC$ the marices computed as the Kronecker products between their respective matrices as defined previously and the Identity matrix $I_3$ (note that the linearized matrices have sizes  ${3\msize}\times{3\ssize}$, ${3\msize}\times{3\csize}$, and ${3\ssize}\times{3\csize}$, respectively). 
Moreover, we explicitly represent the skinning components as follows.
We denote $A \in \mathbb{R}^{3\ssize}$ the vector stacking the current articulations of the skeleton.
And we split the skinning rotational and translational component as follows: we denote $\mathcal{R}$ a ${3\msize}\times{3\msize}$ matrix composed of $\msize$ $3\times{3}$ matrices on the diagonal (block $i$ is the linear part applied to vertex $i$ -- for LBS, $R(i) = \sum_j{\sw{i}{j} \Tri{j}}$), and $T$ the $3\ssize$ vector stacking all translation parameters $\Tti{j}$, where, as previously, $\Tri{j}$ and $\Tti{j}$ are the translation and rotation components of $\Ti{j}$ respectively. 
Therefore, the dynamic rest pose skin $\meshr$, the current skin $\meshc$, and the current cage $\cagec$ are obtained by the following formulas
\begin{equation}
\label{eq_recap_on_the_updates}
\begin{cases}
A = \JC\cdot \cager \\
\tilde{\meshr} = \tilde{\CW} \cdot \cager \\
\tilde{\meshc} = \tilde{\mathcal{R}} \cdot \tilde{\meshr} + \tilde{\SW} \cdot T \\
\cagec = \CWMax^{-1} \cdot \tilde{\meshc} \text{ ,}
\end{cases}
\end{equation}
where used, as in section~\ref{sec_cage_updater}, $\tilde{\cdot}$ to denote all quantities that require only the subset of mesh vertices selected by MaxVol (the cage updater requires the transformed position of those vertices only).
We can observe that the third equation is nothing but a matrix expression of Equation~\ref{eq:skinning}, where the rotational and translational components of the transformation at each vertex have been separated, following Equation~\ref{eq:lbs_with_rotations_and_translations_separated}. 

We aim at computing offsets $\partial\cager$ to apply to the rest cage so as to obtain a resulting offset $\partial\cagec$, which the user wishes to apply to the current cage.
Before pursuing the derivation, we stress that the set of Equations~\ref{eq_recap_on_the_updates} is not sufficient for that purpose as, in fact, applying offsets to the skeleton articulations $A$ results in changes in the skeletal deformation parameters $T$, since the translation parameters are affected to preserve the skeleton connectivity.

\paragraph*{Relating joint offsets and skeletal deformations.}
Following~\cite{thiery2018araplbs}, we note that applying an offset to an articulation $a_j$ results in offsets applied to the translations $\{\Tti{k}\}$ in the following manner
\begin{equation}
\label{eq_equations_to_relate_joint_offsets_and_translations}
\begin{cases}
\Tri{j} \cdot (a_j + \partial{a_j}) + \Tti{j} + \partial{\Tti{j}}  =  a_j + \partial{a_j} & \text{if $j$ is a root} \\
\Tri{j} \cdot \partial{a_j} + \partial\Tti{j} = \Tri{f} \cdot \partial{a_j} + \partial{\Tti{f}} & \text{if $j$ has father $f$} ,
\end{cases}
\end{equation}
the first equation simply means that the pivot point is updated accordingly, and
the second equation simply means that the joint $j$ has to be preserved by the transformations of handle $j$ and its father $f$ both, under preservation of the linear parts $\{\Tri{k}\}$ of the skeletal deformation $\TS$.
This system can be rewritten as
\begin{equation}
\label{eq_relating_joint_offsets_and_translations}
\begin{cases}
(I_3 - \Tri{j}) \cdot \partial{a_j} = \partial{\Tti{j}} & \text{if $j$ is a root} \\
(\Tri{j} - \Tri{f}) \cdot \partial{a_j} = \partial{\Tti{f}} - \partial{\Tti{j}} & \text{if $j$ has father $f$} ,
\end{cases}
\end{equation}
or, in matrix expression
\begin{equation}
\label{eq_relating_joint_offsets_and_translations_matrix_expression}
\mathcal{A}_R \cdot \partial{A} = \mathcal{B}_\text{topo} \cdot \partial{T} .
\end{equation}

We note that, while $\mathcal{A}_R$ depends on the current skeletal deformation parameters (the linear part $\{\Tri{k}\}$), $\mathcal{B}_\text{topo}$ depends on the topology of the skeleton only, and can safely be inverted once and for all, independently of the current skeletal deformation parameter set.

\paragraph{Relating interaction and deformation cages offsets.}
Finally, we can gather the previously derived equations to obtain the $\csize\times\csize$ linear system
\begin{equation}
\label{eq_interaction_and_deformation_cage_offsets}
\left( \tilde{\mathcal{R}} \cdot \tilde{\CW} + \tilde{\SW} \cdot \mathcal{B}_\text{topo}^{-1} \cdot \mathcal{A}_R \cdot \JC \right) \cdot \partial{\cager} = \tilde{\CW}\partial{\cagec}
\end{equation}
that can be resolved efficiently.
Note that when no skeletal deformation is performed i.e., the linear part $\{\Tri{j}\}$ is composed of Identity matrices only and all translations $\{\Tri{j}\}$ are null, the current cage $\cagec$ and rest cage $C$ match (as they should), since $\mathcal{R}$ is then the identity matrix and $\mathcal{A}_R$ is null.

\paragraph{Impact of the MaxVol relaxation}
Note that using a subset of the mesh vertices in the cage fitting has several important consequences:
First, all matrices in Equation~\ref{eq_interaction_and_deformation_cage_offsets} have dimensions bounded by $max(3\csize,3\ssize)$, which results in updates that can be performed in real time on the examples we used, \emph{these timings being in this case entirely independent from the mesh size}.
Secondly, by matching the dimensionality of the cage and the mesh used for inversion, the system in Equation~\ref{eq_interaction_and_deformation_cage_offsets} is exactly invertible, and the loop $\cagec\rightarrow\cager\rightarrow\skelr\rightarrow\skelc\rightarrow\cagec$ is \textbf{exact}.

We originally tried using all vertices in the inversion process, resulting in an approximate loop. While the approximation was extremely subtle and unnoticeable, the biggest issue was that it resulted in reduced performance: we could not obtain results that were fast enough for a modeling session on large models, as the user had to wait a few seconds when switching from skeleton manipulation to cage manipulation.
Note that, if desired, the user could still rely on more vertices than just the ones selected by MaxVol (a good strategy could be to use farthest sampling in the space of cage coordinates of the vertices, as done in~\cite{jacobson2012fast} -- Section~3.3); the construction described in this section would remain valid, but the inverse of the matrices would have to be replaced by pseudo-inverses and the loop would be only approximate (rigorously, the manipulation cage $\cagec$ is then obtained by least-squares fitting as $(\tilde{\CW}^T \cdot \tilde{\CW}) \cdot \cagec = \tilde{\CW}^T \cdot \meshc$, which leads to a modified Equation~\ref{eq_interaction_and_deformation_cage_offsets} where both terms are multiplied by $\tilde{\CW}^T$ on their left).

\paragraph{Handling skeletal deformation methods other than LBS}
As already emphasized earlier, the Reverse Cage Deformer operator assumes LBS as the skinning deformation method.
This will give us indeed an exact result if LBS is the current skinning method, and an approximated result with the other skinning methods. However, since the Reverse Cage Deformer is always applied to small incremental modifications $\partial\cagec$, the approximation error is negligible and hardly noticeable during user interaction. 
Another possibility is to make use of a ghost mesh that is deformed with LBS and use this ghost mesh to drive the fitting of the interaction cage in the cage updater, regardless of the actual final deformation method.
The cage updater {\sc CageUp} is then not optimal, in the sense that it best fits an LBS deformation of the mesh instead of the actual deformed mesh, but the loop $\cagec\rightarrow\cager\rightarrow\skelr\rightarrow\skelc\rightarrow\cagec$ is always exact.
Note that only the vertices selected for inversion need to be deformed in the ghost mesh, so this step is negligible in practice.
Both options are satisfactory and trivial to implement, and will work well for all skinning methods producing deformations that resemble LBS at large scale (as illustrated in our results featuring DQS and CoRs) since the cage fitting performed by {\sc CageUp} is a \emph{global} operation as a result of using global cage weights.

\begin{table*}
\begin{tabular}{|l|ccc|cc|c|c|ccc|}
\hline
\textbf{Model} & \textbf{Verts} & \textbf{Skel}   & \textbf{Cage} & \textbf{SkelUp} & \textbf{CageUp}    & \textbf{CoR}     & \textbf{CageRev}      & \textbf{LBS} & \textbf{DQS} & \textbf{CoR} \\		
         &           & \textbf{joints} & \textbf{handles}	& \textbf{preprocess} & \textbf{preprocess}    & \textbf{update} & \textbf{update} & \textbf{frame}	& \textbf{frame} & \textbf{frame}\\		
&&&& $ms$		& $ms$		& $ms$		& $ms$	& $ms$	& $ms$	& $ms$\\
\hline
Arm (coarse cage) 	& 2089 	& 24 & 28 	& 401 		& 5		& $\leq$1	& 3	& 0.88	& 1.62		& 2.26\\
Arm (medium cage) & 2089 	& 24 & 58 	& 417 		& 29		& $\leq$1 	& 20	& 1.61	& 1.53		& 1.63\\
Arm (fine cage) 		& 2089 	& 24 & 79 	& 439 		& 45		& $\leq$1	& 40	& 2.22	& 1.84		& 2.17\\
Warrok 					& 6557 	& 64 & 104 	& 2672 	& 106	& $\leq$1	& 83	& 7.81	& 7.66		& 9.51\\
Ely 							& 7512 	& 64 & 88 	& 3018 	& 75		& $\leq$1	& 56	& 9.24	& 8.73		& 9.99\\
Airplane 					& 41425 	& 19 & 69 	& 3738	& 44		&  5				& 30	& 23.7	& 21.68	& 26.40\\
Timber Rattlesnake 					& 120066 	& 98 & 44 	& 46771	& 32		&  9				& 16	& 50.03	& 39.43& 52.35\\
\hline
\end{tabular}
\caption{Performances of our modeling system, measured on a MacBook pro early-2015 equipped with an Intel i5 processor, 8GB of RAM and and Intel Iris 6100 GPU. Columns labeled \textbf{SkelUp preprocess} and \textbf{CageUp preprocess} refer to the pre-processing time of the {\sc SkelUp} and {\sc CageUp} operators, respectively. The \textbf{CoR update} column reports the time necessary to update the centers of rotation each time {\sc SkelUp} is executed (this update does not occur when LBS or DQS are used). \textbf{CageRev update} reports the execution time of the {\sc CageRev} operator, when the user switches from skeleton manipulation to cage (in current pose) manipulation during a modeling session. This few-milliseconds latency is observed only when the user grabs the cage and not during cage manipulation after that, as all necessary matrix factorizations do not need being updated as long as the skeleton is untouched. Finally, the last three columns report the cost of updating the current pose with the various skinning methods implemented in the framework (the cost of rendering is not taken into account here). Note that all the timings we report refer to a CPU implementation. Moving to GPU should dramatically improve our performances. Also note that in our examples DQS seems to outperform LBS. While this is not true in the general case, in our codebase we used a carefully optimized implementation of DQS, as opposed to a naive implementation of LBS. Therefore, these numbers are strictly dependent on our specific software prototype.}
\label{tab:res}
\end{table*}

\begin{figure*}[h]
\includegraphics[width=\linewidth]{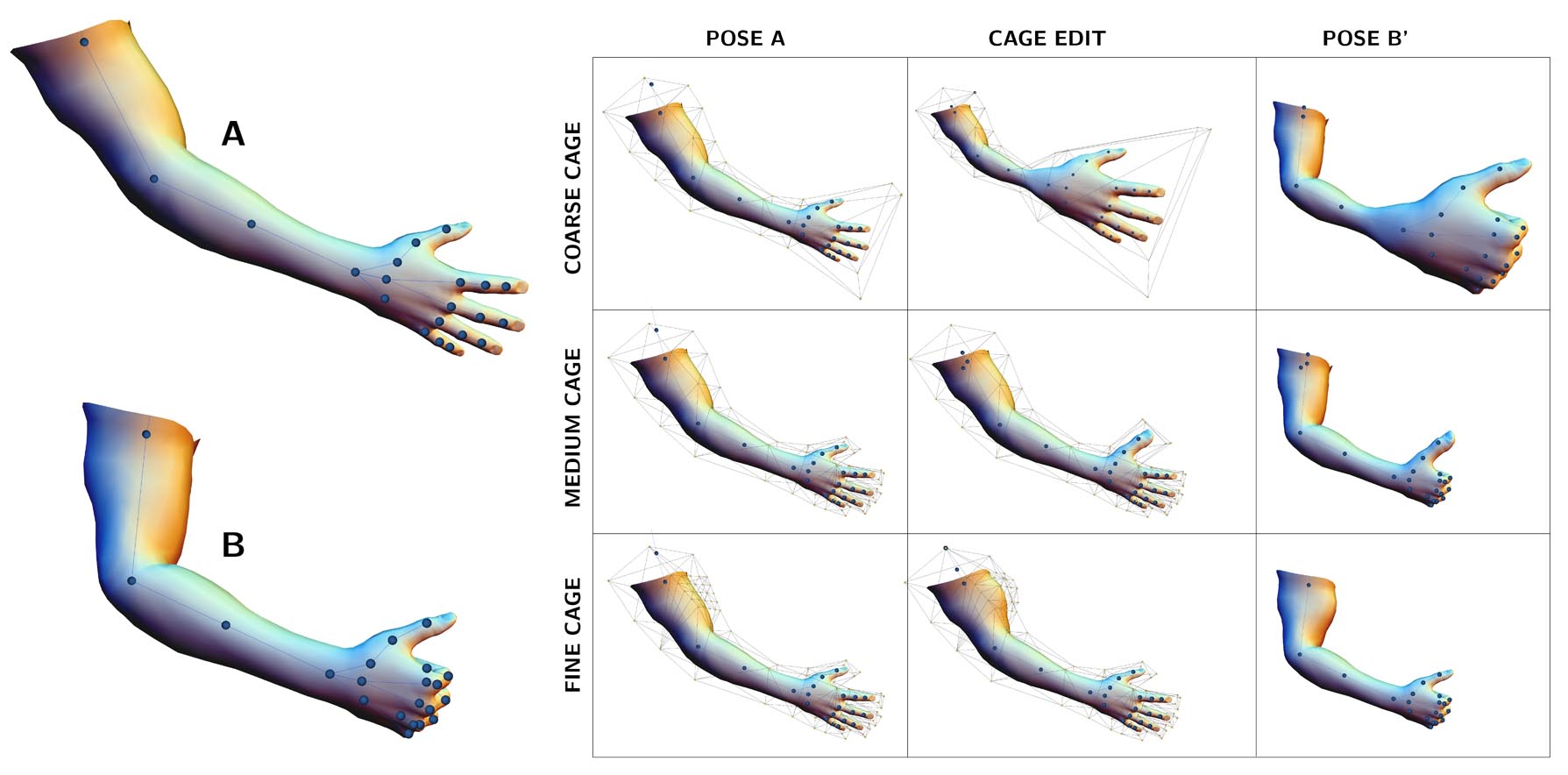}
\caption{Our method can seamlessly combine edits defined on skeletons and cages that operate on features at different scales. Here a simple arm bent with a skeleton (left) is enriched with additional edits with three alternative cages that operate at different levels of details (right). The coarse cage controls the whole hand, and is used to enlarge it; the medium cage allows to selectively edit each finger, and is used to thicken the thumb; the dense cage has many control points around the bicep, and is used to inflate it. All the three cages produce visually plausible deformations that are difficult to replicate by acting solely on the skeleton or on the cage.}
\label{fig:arms}
\end{figure*}

\section{Results}
\label{sec:results}
We have implemented our modeling framework as a single-threaded C++ program. Models have been either manually crafted or downloaded from online repositories such as Adobe Mixamo~\shortcite{mixamo} and SketchFab~\shortcite{sf}. Whenever a control cage was not provided in the original dataset, we used the method proposed in~\cite{CLMASBP19} to produce one. In case the skinning was missing, we manually created one using Maya~\shortcite{maya}.
In terms of performances, our hybrid modeling system introduces only negligible overhead with respect to the classical skeleton- and cage-based pipelines, and for moderately complex characters runs in real time with high frame rate even on commodity hardware (Table~\ref{tab:res}). 

\begin{figure*}
\includegraphics[width=\linewidth]{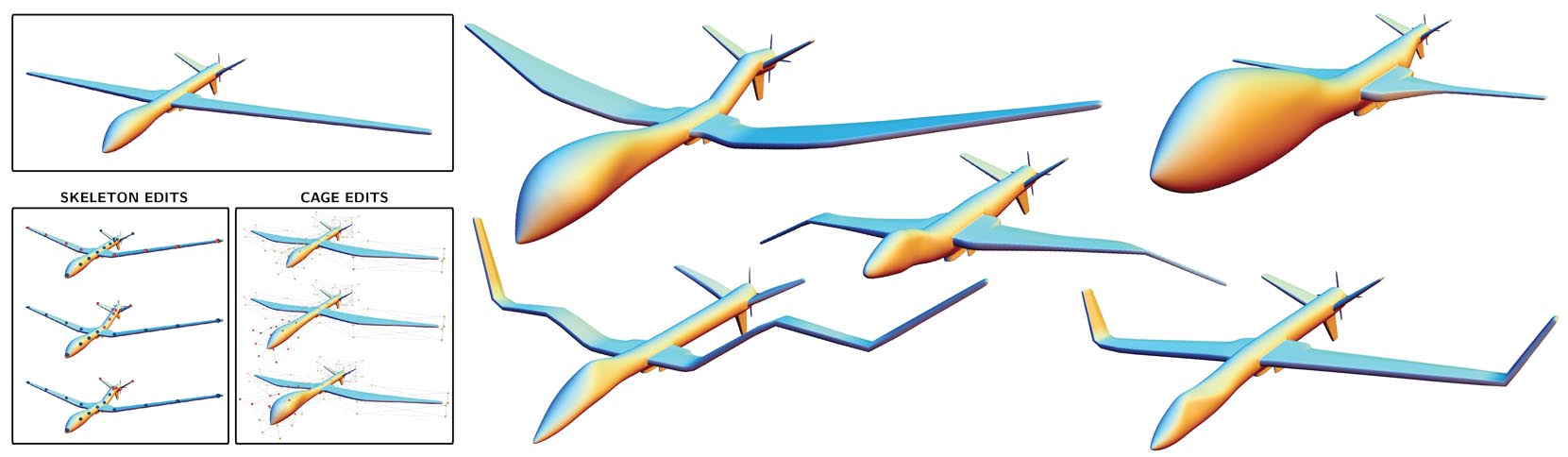}
\caption{Jointly acting on skeletons and cages allows to easily control complementary aspects of the modeling and explore the space of shapes starting from a simple example (top left). Skeletons are best to bend tubular parts and, more in general, deform the rigid components of a shape. Cages are more appropriate to control locally smooth deformations, such as changes of the local thickness of the airplane or the profile of its wings.}
\label{fig:airplane}
\end{figure*}

\begin{figure*}
\includegraphics[width=\linewidth]{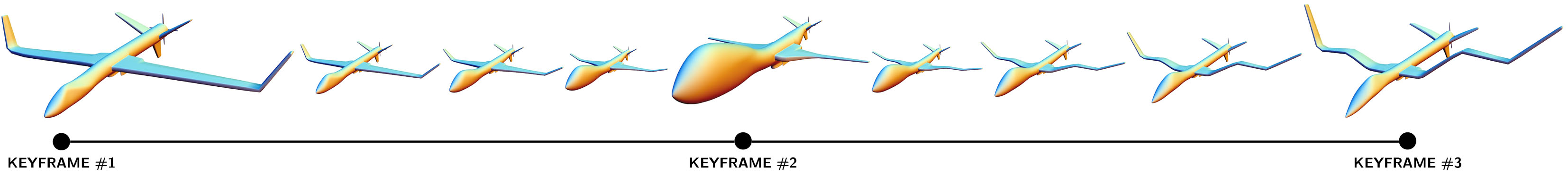}
\caption{Using deformed models as keyframes we can create computer animations. Bone rotations are interpolated using Slerp~\cite{dam1998quaternions}, cage edits are linearly interpolated.}
\label{fig:keyframing}
\end{figure*}

\begin{figure*}
\includegraphics[width=\linewidth]{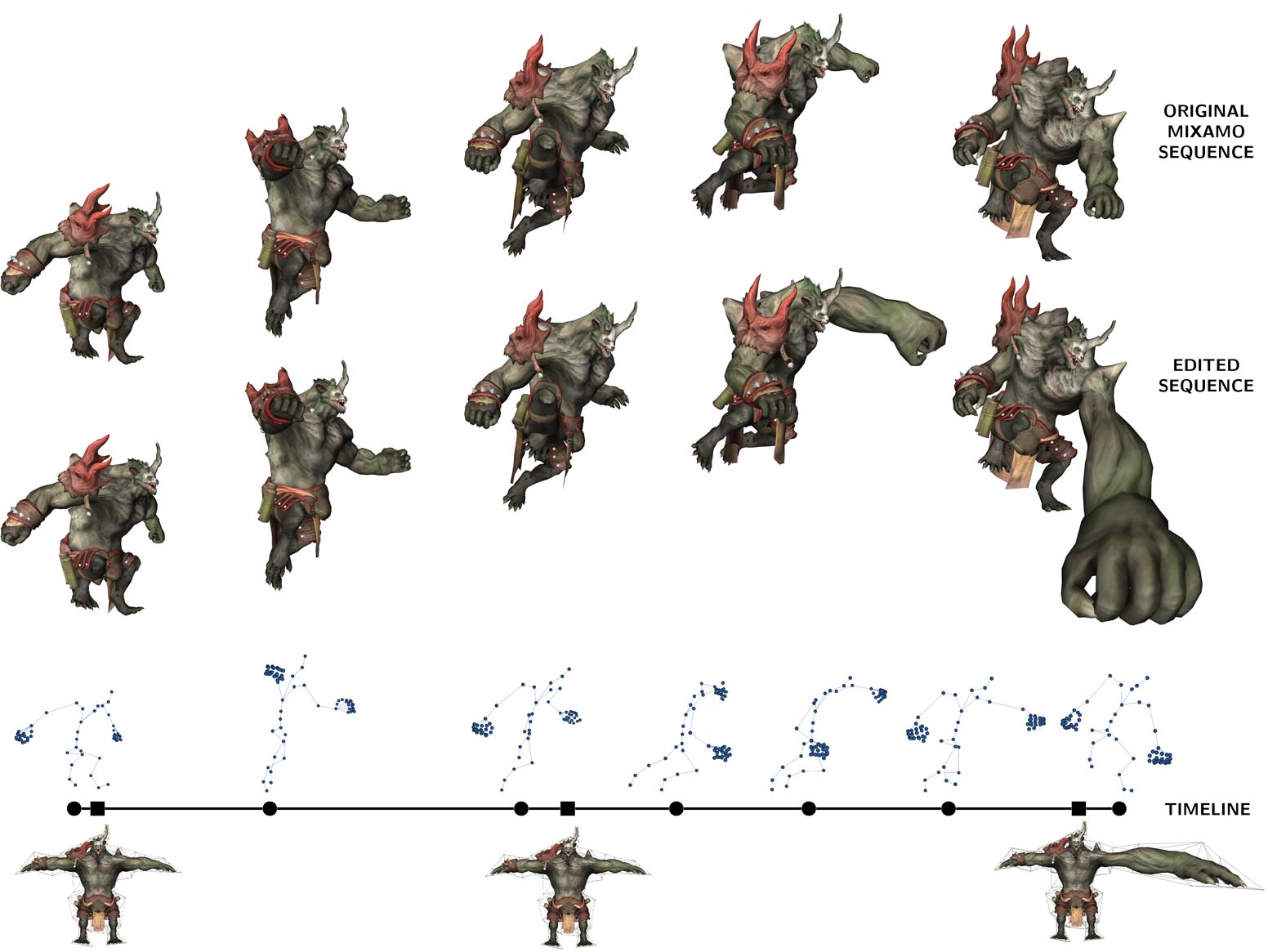}
\caption{Top: a legacy skeleton-based animation downloaded from Adobe Mixamo~\cite{mixamo}. Middle: an edited sequence obtained by inflating the punch with a control cage. Bottom: the animation timeline, with both skeleton keyframes (circles, from Mixamo), and cage keyframes (squares). The first two cage keyframes were automatically generated with~\cite{CLMASBP19}, and simply enclose the rest pose; the third one was manually edited to inflate the punch. Editing a single keyframe we produced a new sequence containing a non trivial twist.
Note also that this example exhibits regions where the skeleton is much more detailed that the cage and vice-versa:
while the hands embed a highly-detailed bone structure allowing animating all the fingers and the cage around them resemble essentially paws, the belly contains a few bones only to mimick a simple spine behaviour while the cage around it is finely detailed to allow for precise anisotropic volume editing.}
\label{fig:warrok}
\end{figure*}

\paragraph*{Deformation options}
A key feature of our hybrid deformation system is its ability to scale across multiple methods for skeleton- and cage-based deformation, which can therefore be chosen from pratictioners depending on their taste and needs. For the skeleton part we implemented the two most popular skeleton-based deformation methods, namely Linear Blend Skinning~\cite{magnenat1989joint}, Dual Quaternions~\cite{kavan2008geometric}, and the recently introduced CoR~\cite{le2016real}, which combines the positive aspects of the previous two and at the same time avoids their weak points (volume loss for LBS, bulging for DQS). A side by side comparison between these three alternatives is shown in Figure~\ref{fig:compSkinning}.
For the skinning weights, although various automatic methods exist in literature (Section~\ref{sec:related}), industry level deformations often involve carefully designed weights that are manually painted on the surface by skilled artists. Our system is agnostic on the specific weights of choice, and transparently supports both automatic and manual approaches. Rigs are imported into the system using standard formats (i.e. FBX),
securing an easy interface with commercial software and publicly available repositories.
For the cage part we used the Mean Value Coordinates~\cite{ju2005mean} in all our tests, which are internally computed by our framework. 
Similarly to skeleton weights, alternative barycentric coordinates that obey the linear blend of Equation~\ref{eq:caging} can be 
loaded into the system and used in a transparent way. To the best of our knowledge, this includes the vast majority of the known barycentric coordinates that appeared in literature, including the recently proposed coordinates for quad cages~\cite{TMB:2018:QMVC}. Two notable exceptions are the Green Coordinates~\cite{lipman2008green} and the Variational Harmonic Maps~\cite{ben2009variational}, which both use a blend equation that involves mesh vertices and face normals, and are therefore not directly applicable to our linear deformation paradigm.


\paragraph*{Skeleton updater} In Figure~\ref{fig:eval_skelupdater} we evaluate our skeleton updater with a synthetic shape, consisting on a straight bar bent at 45/90/135 degrees using LBS. Editing the bar with the cage moves the skeleton away from the correct centers of rotation, and without the skeleton re-fitting procedure described in Section~\ref{sec:supd} extremely evident artifacts arise. Our bone positioning system correctly recovers from all configurations, producing visually plausible deformations. Note that the system is not sensitive to the specific cage being used, and produces valid results both with a regular (left) and irregular (right) cage.

\paragraph*{Scale adaptivity}
Skeletons and cages may be very dfferent to one another, and are indeed able to control features of the same object at different scales. Our system is able to seamlessly combine skeletons and cages that operate at different levels of detail. In Figure~\ref{fig:arms} a simple bent arm controlled by a skeleton is further edited with three alternative cages. The coarse cage controls the whole hand, and is used to enlarge it; the medium cage allows to selectively edit each finger, and is used to thicken the thumb; the dense cage contains various control points around the bicep, and is used to inflate it when the arm is bent. All three hybrid deformations are visually plausible and difficult to replicate by acting solely on a skeleton or on a cage.

\paragraph*{Hybrid modeling} The principal intent when designing our deformation framework was to offer artists a unique system where they could seamlessly combine multiple deformation paradigms. Starting from an input shape linked to a skeleton and a cage, artists can explore the space of deformations to create a family of similar objects, using the more appropriate tool for each edit. An example of hybrid modeling is given in Figure~\ref{fig:airplane}, where several variations of an airplane are produced from a single item. Skeleton bones are used to control the rigid parts the plane (e.g. to bend the core and the wings). Cage handles are used to apply local volumetric deformations, for example to locally inflate parts of the core, or to edit the profile of the wings.

\begin{figure}[h]
  \centering
  \includegraphics[width=0.99\linewidth]{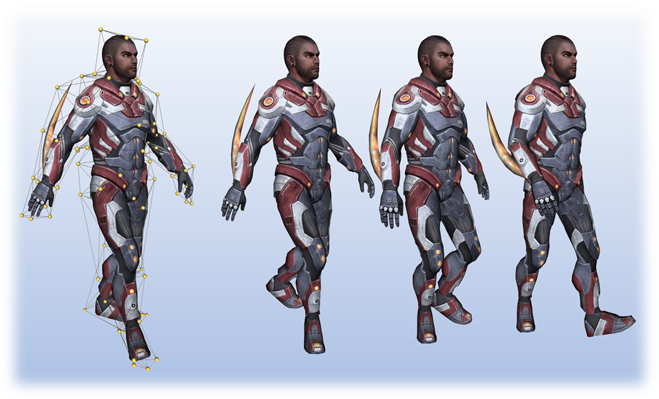}\\
  \includegraphics[width=0.99\linewidth]{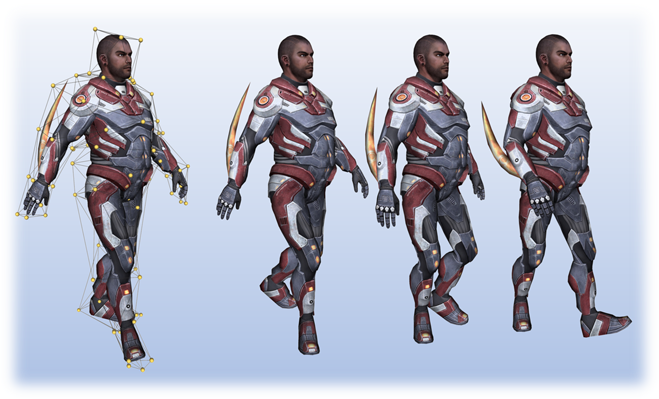}
  \caption{Our framework allows animating and deforming characters simultaneously. In the top row we show the Mixamo's Ely character, with a cage we added on the left, and three frames of the walking sequence. In the bottom row we deformed the same character fattening him (notice the changed cage on the left), and performed the same walking animation. Combining skeleton and cage controls, we can fatten the character while it walks as we show in the accompanying video.}
  \label{fig:walkingChar}
\end{figure}

\paragraph*{Hybrid Animation}
Another interesting application of our framework consists in using the various shapes it produces as keyframes, to guide a computer generated animation sequence. In Figure~\ref{fig:keyframing} we show a few interpolated frames of an animation, obtained keyframing some of the airplanes shown in Figure~\ref{fig:airplane}. In between frames are generated interpolating bone rotations with Slerp~\cite{dam1998quaternions} and cage vertices linearly. Note that the \emph{deformation} cage $\cager$ is keyframed, not the manipulation cage $\cagec$. The skeleton updater is therefore required at each reconstruction step (but the update is extremely fast as it is linear in the number of skeleton joints only), but the cage reverse updater is not involved in the process.
Following a similar approach legacy animations can be enriched with new effects. In Figure~\ref{fig:warrok} we show a skeleton driven punch sequence downloaded from Adobe Mixamo~\cite{mixamo}, which we enriched with three cage keyframes that inflate the punch at the proper time. Note that a minimal workload is already enough to incorporate in the animation new interesting effects. In this specific case only one manually edited keyframe was used, the other two are simple envelopes of the rest pose, computed with~\cite{CLMASBP19}. Also note that skeletons and cage keyframes are interpolated asynchronously, hence can work on the same character independently and at different levels of detail.
Similar results are also shown in Figure~\ref{fig:walkingChar}.
We point the reader to the accompanying video to see the actual animations we obtained.

%
%
%




\section{Conclusion and future work}
\label{sec:conc}
We started from the observation that skeleton-based and cage-based deformations control different aspects of shape modeling, and are to a large extent complementary to one another. We therefore proposed a real-time modeling framework based on a novel paradigm that seamlessly combines these structures. We obtained the desired effect by adopting the concept of rest pose and current pose for both skeletons and cages, introducing novel update operators that realize the sync between all these structures.
As a result, we operate in a larger deformation space, containing
poses that are impossible to obtain by acting solely on a skeleton or a cage.

Our framework is back-compatible with most existing techniques for skeleton-based and cage-based deformation.
The only limitation in this sense comes from our assumption of a linear equation for cage-based shape editing (Equation~\ref{eq:caging}). From a technical standpoint, non linear cage-based deformation techniques such as~\cite{lipman2008green} could potentially be incorporated in the system, though at the cost of having more complex algorithms to maintain the sync. Similar considerations can be done for partial cages, such as the ones proposed in~\cite{Garcia:2013:CMM}. We did not perform tests in these directions yet, but it would be interesting to check how this will affect the frame rate and the real-time experience. 

Regarding user interaction, the visualization of controllers for both the skeleton and the cage may sometimes clutter the screen, especially for complex characters requiring numerous skeleton bones and complex control cages. We envisage a potential improvement by adopting a dynamic rendering of the controllers, that fades away from the mouse position.

We believe that our contribution 
could support advanced deformation control.
In our future work, we plan to extend the framework with more controllers (e.g. point handles) which we can incorporate with the same approach to synchronization, while remaining oblivious on the specific technique used for their implementation.





\bibliographystyle{ACM-Reference-Format}
\bibliography{ms}

\end{document}